\documentclass[10pt]{article}

\usepackage{a4wide}
\usepackage{mediabb}

\usepackage{epsf}
\usepackage{color}
\usepackage{amsmath}
\usepackage{amssymb}
\usepackage{latexsym}
\usepackage{bm}
\usepackage[T1]{fontenc}
\usepackage{slashed}

\DeclareMathOperator{\diag}{diag}

\newcommand{\A}{\mathcal{A}}

\newcommand{\ul}{\ensuremath{\underline}}
\newcommand{\ol}{\ensuremath{\overline}}

\newcommand{\al}[1]{\begin{align}#1\end{align}}
\newcommand{\als}[1]{\begin{align*}#1\end{align*}}
\newcommand{\bp}{\begin{pmatrix}}
\newcommand{\ep}{\end{pmatrix}}
\newcommand{\nn}{\nonumber\\}
\newcommand{\paren}[1]{\left(#1\right)}

\newcommand{\mc}{\mathcal}

\newcommand{\del}{\partial}
\newcommand{\edh}{\text{\dh}}
\newcommand{\edth}{\text{\dh}}
\newcommand{\edthbar}{\bar{\text{\dh}}}
\newcommand{\edhb}{\bar{\text{{\dh}}}}

\newcommand{\red}{\textcolor{black}}

\newcommand{\ns}{N\atop S}
\newcommand{\sn}{S\atop N}

\begin{document}
\title{
Universal Extra Dimensions on Real Projective Plane\\
\medskip
}
\author{
	\Large Hideto Dohi\thanks{E-mail: \tt dohi@het.phys.sci.osaka-u.ac.jp
	}
	\ and
	Kin-ya Oda\thanks{E-mail: \tt odakin@phys.sci.osaka-u.ac.jp}\medskip\\
	\it Department of Physics,  
	Osaka University,  Osaka 560-0043, Japan\\ \medskip
	}

\maketitle
\begin{abstract}

\noindent We propose a six dimensional Universal Extra Dimensions (UED) model compactified on a real projective plane $RP^2$, a two-sphere with its antipodal points being identified. We utilize the Randjbar-Daemi-Salam-Strathdee spontaneous sphere compactification with a monopole configuration of an extra $U(1)_X$ gauge field
that leads to a spontaneous radius stabilization.
Unlike the sphere and the {\color{black} $S^2/Z_2$ orbifold} compactifications, the massless $U(1)_X$ gauge boson is safely projected out.
We show how a compactification on a non-orientable manifold results in a chiral four dimensional gauge theory by utilizing 6D chiral gauge and Yukawa interactions.
The resultant Kaluza-Klein mass spectra are distinct from the ordinary UED models compactified on torus.
We briefly comment on the anomaly cancellation
and also on a possible dark matter candidate in our model.


 \end{abstract}
\vfill
\hfill OU-HET-664
\newpage

\section{Introduction and Conclusions}
\label{Introduction_and_Conclusions}
Universal Extra Dimensions (UED) allow fairly lower Kaluza-Klein (KK) scale even after taking into account the precision electroweak constraints~\cite{Appelquist:2000nn,Appelquist:2002wb,Gogoladze:2006br}, comparing to a model where some of the Standard Model (SM) fields are localized in the extra dimensions, and therefore the UED can be directly tested at the CERN Large Hadron Collider (LHC) which is just being started.
Furthermore the Lightest KK Particle (LKP) in the UED model can serve as a natural dark matter candidate~\cite{Cheng:2002iz,Servant:2002aq}.
A six dimensional (6D) UED model is particularly attractive since three generations of SM fermion are naturally required~\cite{Dobrescu:2001ae,Borghini:2001sa} for the cancellation of $SU(2)_W$ global anomaly, which is analogous to the Witten anomaly.\footnote{\label{anomaly_footnote}
This three-generation argument of global anomaly is based on the computation with $S^6$~\cite{Bershadsky:1997sb}. Strictly speaking, when extra dimensions are compactified on $B$, analysis must be redone on $S^4\times B$.
We thank S.\ Yamaguchi for pointing out this issue.
}

In 1953, Pauli initiated the compactification on two-sphere $S^2$~\cite{O'Raifeartaigh:1998pk,Straumann:2000zc} and
DeWitt further developed it in 1963 (see~\cite{Pons:2006vz} for a historical review).
After a half century, it has been still attracting interest as a beautiful tool to incorporate the spontaneous compactification (see~\cite{RandjbarDaemi:1982hi} and citations thereof), the gauge-Higgs unification~\cite{Manton:1979kb,Hosotani:1983vn,Hatanaka:1998yp,Dvali:2001qr,Lim:2006bx,Nomura:2008sx}, supergravity models~\cite{Salam:1984cj,Fujii:1986nh} (see also~\cite{Salvio:2007mb} and references therein), etc.
Recently, a UED model is constructed on the {\color{black} $S^2/Z_2$ orbifold}~\cite{Maru:2009wu}, in which there exists a stable particle due to the ``$Z_2'$ KK parity,'' potentially being a possible dark matter candidate.

In this paper we first consider a UED model compactified on the two-sphere, based on the spontaneous compactification mechanism by Randjbar-Daemi, Salam and Strathdee (RSS), which utilizes a monopole configuration of an extra $U(1)_X$ gauge field in the two-sphere~\cite{RandjbarDaemi:1982hi}.
A remarkable feature of the RSS sphere compactification is that, in contrast to the ordinary UED models on torus, the compactification radius is spontaneously fixed by the monopole configuration.\footnote{
See Refs.~\cite{Okada:1984cv,Okada:1984sf} for cosmology of the RSS model.
}
We show that our model can supply a possible dark matter candidate, due to what we call the KK angular momentum.
Actually the stable particle is exactly the same one as in~\cite{Maru:2009wu} and the so-called $Z_2'$ parity conservation is nothing but the result of the KK angular momentum conservation.\footnote{
We also note that our two-sphere compactification can serve chiral fermions~\cite{RandjbarDaemi:1982hi} so that the $Z_2$ orbifold projection in~\cite{Maru:2009wu}, which is $(\theta,\phi)\sim(\pi-\theta,-\phi)$ in the standard polar coordinates, is rather redundant.
}

However, the two-sphere compactification suffers from the existence of massless $U(1)_X$ gauge boson, which obviously contradicts existing experimental observation~\cite{Smith:1999cr} (see also~\cite{Adelberger:2009zz} for a review).\footnote{\label{anomaly_giving_superheavy}
In the {\color{black} $S^2/Z_2$ orbifold} model~\cite{Maru:2009wu}, there exists exactly the same problem and it is assumed that {\color{black} the SM-singlet $\mc N_-$ is not put and hence} the $U(1)_X$ is broken by an anomaly~\cite{Scrucca:2003ra}.
It is somewhat puzzling how a classical gauge configuration is allowed when $U(1)_X$ is broken by an anomaly at an UltraViolet (UV) scale $\Lambda$.
That is, if we have a gauge boson mass term like $\Lambda^2X_MX^M$~\cite{Scrucca:2003ra}, it will spoil the monopole vacuum expectation value (vev)~\cite{RandjbarDaemi:1982hi} due to the extra contribution to the energy momentum tensor in the Einstein equation.
}
As a remedy, we propose a compactification on a real projective plane $RP^2$ that is a two-sphere with its antipodal points being identified: $(\theta,\phi)\sim(\pi-\theta,\phi+\pi)$.\footnote{
A compactification on a flat projective plane is considered in a different context in~\cite{Hebecker:2003we}, where all the curvature is put into conical singularities. {\color{black} This singular space is also utilized to construct another UED model~\cite{Cacciapaglia:2009pa}.}
}
By this identification, the unwanted zero mode of $U(1)_X$ gauge field is projected out, as well as the first KK photon that is a possible dark matter candidate in the $S^2$ and $S^2/Z_2$ UED models.
In our $RP^2$ model, $U(1)_X$ gauge interaction is made anomaly free and the first KK mode of $U(1)_X$ gauge field becomes a possible dark matter candidate.\footnote{
In 5D UED model, radiative corrections tend to make gauge bosons lighter than fermions~\cite{Cheng:2002iz}. If this turns out to be the case in our model, the $U(1)_X$ KK gauge boson becomes the dark matter. Generically one expects that there are also tree level threshold corrections at UV cut-off scale. In that case, the first KK neutrino might become the dark matter.
}
We also discuss possible cosmological constraint from the decay of the second lightest KK particle into the dark matter.

The $RP^2$ is a non-orientable manifold and so is the whole space $M^4\times RP^2$, where $M^4$ is the four dimensional (4D) Minkowski space.
It is standard lore that a quantum field theory must be either $P$ or $CP$ symmetric on a non-orientable manifold and hence cannot account for the SM~\cite{Zeldovitch_Novikov,Ellis:1970ey}.
This argument is based on the fact that a non-orientable manifold necessarily has a closed loop that gives one a parity transformation $P$ when one circulates around it.
Therefore, the Lagrangian must be symmetric under either $P$ or $CP$ transformation for the theory to be single-valued on the space.
In four dimensions, this is obviously in contradiction to the experimentally observed $P$ and $CP$ violation.

The key observation is that the higher dimensional $CP$ transformation is different from the four dimensional one.
We can identify the field after the circulation of the closed loop by an outer automorphism~\cite{Hebecker:2001jb} with the 6D-$CP$ transformation.
{\color{black} Even if a} theory is vector-like (non-chiral) under the 6D-$CP$ transformation, it can be chiral under the 4D-$P$ and $CP$ transformations after the compactification.
{\color{black} The (4D) $CP$ violation in Yukawa sector results} from the fact that we can identify SM spinors by 6D-$CP$ transformation and Higgs by 6D-$P$ transformation.

We obtain the KK mass spectrum that is distinctive from those in the 5D and 6D UED models with orbifolded toroidal compacitifications: $S^1/Z_2$ and $T^2/(Z_2\times Z_2')$, respectively.
From the conservation of the KK angular momentum, the single production of a KK mode from SM matters cannot occur, in contrast to the ordinary UED where only KK parity is conserved and hence one can produce a single KK-even mode from a scattering of zero modes.
Our model contains a possible dark matter candidate, the first KK mode of the $U(1)_X$ gauge field, as well as the more conventional one, the fist KK neutrino, depending on the  input at the UV cutoff.

\section{RSS Spontaneous Compactification on Sphere}
\label{RSS_review}
First let us review  the RSS spontaneous compactification on sphere~\cite{RandjbarDaemi:1982hi}. 
We put the following $M^4\times S^2$ ansatz on the metric
\al{
ds^2 &= \eta_{\mu\nu}dx^\mu\,dx^\nu+R^2\paren{d\theta^2+\sin^2\theta\,d\phi^2},
}
where Greek indices $\mu,\nu,\dots$ run for uncompactified dimensions $0,1,2,3$ and the constant $R$ is the compactification radius. 
This metric does not satisfy Einstein equation with {\color{black} any} six dimensional cosmological constant $\Lambda_6$. 
To satisfy the Einstein equation, we introduce an extra $U(1)_{X}$ gauge field $X:=X_Mdx^M$ with a classical monopole configuration:
\al{
X{}^{\ns}&=\frac{n}{2g_{X}}(\cos\theta \mp1)d\phi,  
	\label{monopole_classical_configuration}
}
where
	$g_X$ is the gauge coupling constant,
	upper-case indices $M,N,\dots$ run for $0,1,2,3,\theta,\phi$, and
	the superscripts $N$ and $S$ denote that the field is written in the north and south patches, respectively.\footnote{
If we simply put $X\propto\cos\theta\,d\phi$, the Einstein equation is satisfied but the configuration becomes singular at the poles.
}
From the single-valuedness of the transition function between two patches, the monopole charge $n$ is imposed to be an integer, which we take $n>0$ without loss of generality. 
The Einstein equation fixes the radion, that is, the compactification radius $R$ is fixed to be:
\al{
R&=\frac{\kappa n}{2g_X},\label{radius}
}
where $\kappa$ is the six-dimensional gravitational coupling constant ($\kappa=\sqrt{8\pi G_6}$).\footnote{
In order to impose the uncompactified space to be flat, we need to fine-tune the six dimensional cosmological constant to be: $\Lambda_6=2g_{X}^2/\kappa^2 n^2$. Here we do not seek for a solution to the cosmological constant problem.
}
This is the RSS spontaneous compactification on sphere.

When the compactification scale is accessible for the LHC: $R^{-1}\sim\text{TeV}$, 
we find that the 6D gravitational constant\footnote{
We note that the 6D gravitational scale must be generally of order $\sim 10^7\,\text{TeV}$ in any 6D UED model.
If there are further extra dimensions in which only gravity propagates, the higher dimensional gravitational scale needs not to be the same as the 4D Planck scale but can be lower than that, even around TeV for large~\cite{ArkaniHamed:1998rs} and warped~\cite{Randall:1999ee} extra dimensions.
}
is required to be of order $\kappa^{-1/2}\sim 10^7\,\text{TeV}$ and resultantly that the effective four-dimensional gauge coupling $g_{X4}:=g_X/\sqrt{4\pi}R$ becomes: $g_{X4}\sim n\times10^{-14}$.
Note that a 6D gauge theory is necessarily a non-renormalizable effective theory with a UV cutoff scale $\Lambda$ and that the effective theory is valid when $g_X\ll\Lambda^{-1}$, that is, when $\Lambda\ll 10^{14}\,\text{TeV}$ for $R^{-1}\sim\text{TeV}$. This condition is well satisfied for the 6D UED model since it must be that $\Lambda\ll \kappa^{-1/2}\sim 10^7\,\text{TeV}$.
This set of parameters $\kappa$ and $g_{X4}/n$ is also used in~\cite{Maru:2009wu} implicitly.


Now that the background spacetime is fixed, let us consider the KK reduction of bulk fields with spin zero, half and one.
Hereafter we neglect the gravitational fluctuations.
In~\cite{RandjbarDaemi:1982hi}, KK wave function is given as a harmonic expansion in terms of the Wigner $D$-matrix.
Here we reformulate it in terms of the Newman-Penrose $\edth$ formalism and the spin-weighted spherical harmonics~\cite{Newman:1966ub,Goldberg:1966uu,Castillo}.
The $\edth_s$ and $\edthbar_s$ operators, acting on an object $\eta_s$ with spin-weight $s$, are defined by\footnote{
Note that the complex conjugate of $\edth_s$ becomes $\ol{\edth_s}=\edthbar_{-s}$; the left-hand-side notation will not be used throughout this Letter. Conventionally, the subscript $s$ is omitted from $\edth_s$ but we put it for explicitness.
}
\al{
\edh_s\,\eta_{s}:
=& -\paren{\del_\theta+i\csc\theta\del_{\phi}-s\cot\theta}\eta_{s},\label{edh_body}
\\
\edhb_s\,\eta_{s}:
=& -\paren{\del_{\theta}-i\csc\theta\del_{\phi}+s\cot\theta}\eta_{s}.\label{edhbar_body}
}
A spin-weighted spherical harmonics ${}_sY_{jm}$ has spin-weight $s$ and is defined iteratively by
\al{
\edh_s \,{}_{s}Y_{jm}=&\sqrt{j(j+1)-s(s+1)}\,{}_{s+1}Y_{jm},\label{sYjm000}\\
\edhb_s \,{}_{s}Y_{jm}=&-\sqrt{j(j+1)-s(s-1)}\,{}_{s-1}Y_{jm},\label{sYjm}
}
from the ordinary spherical harmonics ${}_0Y_{jm}:=Y_{jm}$.
Note that the $\edth$ and $\edthbar$ operators increase and decrease spin weight by $+1$ and $-1$, respectively.
The spin-weighted spherical harmonics is the eigenfunction of the operator
\al{
\textbf{K}_s^{2}
	&:=	-\edthbar_{s+1}\,\edth_s+s(s+1)
	=		-\edth_{s-1}\,\edthbar_s+s(s-1),
}
with the eigenvalue
\al{
\textbf{K}_s^{2}\,{}_sY_{jm}
	&=	j(j+1)\,{}_sY_{jm}.
}

A bulk scalar field with $U(1)_X$ charge $q$ has the following transition function between north and south patches:
\al{
{\Phi(x,\theta,\phi)}^S
	=	e^{-2is\phi}{\Phi(x,\theta,\phi)}^N,
		\label{scalar_transition}
}
where the spin-weight takes the value $s=qn/2$ for a scalar field.
The transition function~\eqref{scalar_transition} is single-valued for $\phi\to\phi+2\pi$ when $qn$ is an integer.
The KK expansion is given as:
\al{
{\Phi(x,\theta,\phi)}^{\ns}
	&=	\sum_{j=|s|}^\infty\sum_{m=-j}^j\varphi^{jm}(x)\,
		{f_\Phi^{jm}(\theta,\phi)}^{\ns}, &
{f_\Phi^{jm}(\theta,\phi)}^{\ns}
	&:=	{{}_sY_{jm}(\theta,\phi)e^{\pm is\phi}\over R},
	\label{KK_wave_function_for_scalar}
}
where ${}_sY_{jm}$ is the spin-weighted spherical harmonics.
Note that although the KK wave function~\eqref{KK_wave_function_for_scalar} is given for each patch with the transition~\eqref{scalar_transition}, the 4D modes $\varphi^{jm}(x)$ are independent of patches as it must be.
The KK mass for a $jm$-mode is given independently of $m$ by
\al{
M_j	&=	\sqrt{{j(j+1)-s^2\over R^2}+M^2},
}
where $M$ is the bulk mass.
In particular, the lowest $j=|s|$ mode mass becomes $M_{|s|}=\sqrt{{|s|\over R^2}+M^2}$.
Number of (real) degrees of freedom is $2(2j+1)$ for a $j$-th mode.
In the following, we will take the bulk Higgs to be neutral under $U(1)_X$: $q=0$.

Next we consider KK expansion of a bulk spinor field with $U(1)_X$ charge $q$. For the spin connection to be regular at north and south poles, it is convenient to chooze the following vielbein basis, defined on north and south patches, respectively:\footnote{
Our vielbein is related to that in~\cite{RandjbarDaemi:1982hi} by $e^{\ul{5}}=-E^5$ and $e^{\ul{6}}=-E^6$.
}
\al{
\bp e^{\ul{5}}\\ e^{\ul{6}}\ep^{\ns}&:=\bp \cos\phi & \pm \sin\phi \\ \mp \sin\phi & \cos\phi \ep \bp e^{\ul{\phi}} \\ e^{\ul{\theta}} \ep, \label{vielbein_basis}
}
where $e^{\ul\theta}$ and $e^{\ul\phi}$ are vielbein 1-forms with $e^{\ul N}:=e_M{}^{\ul N}dx^M$ and
\al{
e_{M}{}^{\ul{N}}&:=\diag{\paren{1,1,1,1,R,R\sin\theta}}.
}
Here and hereafter, underlined indices denote the (co)tangent space coordinates.
We follow the convention in~\cite{RandjbarDaemi:1982hi,DO_to_appear} for the tangent space gamma matrices $\Gamma^{\ul A}$ ($A=0,1,2,3; 5,6$), then the resultant spin connection 1-form reads
\al{
{\Omega}^{\ns}
	=\paren{\cos\theta\mp1}\Sigma^{\ul{5}\ul{6}}\,d\phi
	={i\over2}\paren{\cos\theta\mp1}\bp\gamma^{\ul 5}\\ &-\gamma^{\ul 5}\ep d\phi,
}
where $\Sigma^{\ul{A}\ul{B}}:={1\over4}\left[\Gamma^{\ul A},\Gamma^{\ul B}\right]$ are generators of local Lorentz transformations.
We decompose the six-dimensional eight component spinor into the four and two component ones $\psi$ and $\chi$, respectively:
\al{
\Psi
	=	\bp\psi_+\\ \psi_-\ep
	=	\bp\chi_{+L}\\ \chi_{+R}\\ \chi_{-L}\\ \chi_{-R}\ep,
}
where $\pm$ denotes the 6D-chirality (eigenvalues of $\Gamma^{\ul 7}$) and $L$ and $R$ denote the usual 4D-chirality, with the $\gamma^{\ul 5}$ eigenvalues $+1$ and $-1$, respectively.\footnote{
The 4D-left and right chiralities are defined differently in~\cite{RandjbarDaemi:1982hi}.
}
Note that each of six-dimensional Weyl fermion $\Psi_\pm:={1\pm\Gamma^{\ul 7}\over2}\Psi$ forms an irreducible representation of the six-dimensional local Lorentz group.

Thanks to the basis~\eqref{vielbein_basis}, the spin connection takes the same form as the background monopole configuration~\eqref{monopole_classical_configuration}.
Resultantly, the bulk spinor fields given in north and south patches in the overlap region $0<\theta<\pi$ are connected to each other by the local Lorentz and $U(1)_X$ transformation:
\al{
{\Psi(x,\theta,\phi)}^S
	=
		e^{-2\phi\Sigma^{\ul{5}\ul{6}}}
		e^{-iqn\phi}
		{\Psi(x,\theta,\phi)}^N.
		\label{NS_neutral_spinor_transition}
}
The transition function~\eqref{NS_neutral_spinor_transition} is single-valued for $\phi\to\phi+2\pi$ when $qn$ is an integer.
In the model, we will assign the $U(1)_X$-charge to be $qn=\pm1$ for the SM bulk fermions.

The KK expansion is most conveniently given in terms of the two component spinors~\cite{DO_to_appear}:
\al{
{\chi_\alpha(x,\theta,\phi)}^{\ns}
	&=	\sum_{j=|s|}^\infty\sum_{m=-j}^j
		\chi_\alpha^{jm}(x)
		f_\alpha^{jm}(\theta,\phi)^{\ns}, &
f_\alpha^{jm}(\theta,\phi)^{\ns}
	&=	{{}_sY_{jm}(\theta,\phi)e^{\pm is\phi}\over R},
}
where $\alpha$ stands for $+L$, $+R$, $-L$ and $-R$
and
\al{
s={qn\pm1\over2} \quad\text{for}\quad \alpha=\begin{cases}+L, -R,\\ +R, -L,\end{cases}
}
with $\pm$ signs corresponding to the $\alpha$-cases above and below.
The KK mass for a $jm$-mode is given independently of $m$ as
$\sqrt{{j(j+1)\over R^2}-{(qn)^2-1\over4R^2}+M^2}$, where $M$ is a possible 6D-Dirac mass term $M\ol{\Psi_+}\Psi_-+\text{h.c.}$ that is allowed when $\Psi_+$ and $\Psi_-$ have the same $U(1)_X$ charges (as well as all the same unbroken gauge charges).\footnote{
Unlike the model with position dependent mass~\cite{Park:2009cs}, the presence of a constant bulk mass term does not lead to a modification of the wave function profiles in extra dimensions.
In particular, the zero mode wave function profile is kept flat when $qn=\pm1$ even if one turns on $M$.
This is not the case when we have a warped (un-factorizable) metric.
}

We note that the zero mode is the $j=j_0:={|qn|-1\over2}$ mode, which becomes 4D-chiral so that there exists only $L$ or $R$ component in it.\footnote{
When the bulk fermion does not have a $U(1)_X$ charge: $q=0$, the lightest $\ell=1$ mode is massive which has both 4D-chiralities $L$ and $R$.
}
Let us then consider a $\ell$-th massive KK mode $j=j_0+\ell$ with $\ell\geq1$.
This mode is 4D vector-like and its 4D-Dirac mass (between fermions in mass eigenbasis) is
\al{
M_\ell=\sqrt{{\ell\left(\ell+|qn|\right)\over R^2}+M^2}.
	\label{spinor_KK_masses}
}
Per each 6D-chiral field, number of degrees of freedom is $2(2j_0+1)$ for a (4D-chiral) zero mode and $4[2(j_0+\ell)+1]$ for each (4D-vector-like) $\ell$ mode.
We will choose the $U(1)_X$ charge of the SM spinors to be either $qn=\pm 1$.
In that case, relative number of degrees of freedom to a $L$ or $R$ zero mode is $2\ell+1$ for each $L$ and $R$ KK mode.

\red{Next} we show the KK expansion for a bulk vector field $\A:=\A_Mdx^M$.
Although $U(1)_{X}$ gauge field has a classical configuration, KK expansion of its quantum fluctuation is the same as the other SM gauge fields since the $U(1)_X$ gauge field is neutral under itself.
The SM gauge fields are obviously neutral under $U(1)_X$ and are not affected by the monopole configuration.

Let us define the following complex fields as combinations of local Lorentz vectors:\footnote{
These fields can be written in terms of a complex adjoint 6D scalar $\Phi_{\A}$:
\als{
\A_{+}&=-i\edh_0\, \Phi^{\dagger}_{\A}, & \A_{-}&=i\edhb_0\, \Phi_{\A}.
}  
}
\al{
\A_{\pm}&:=\frac{1}{\sqrt{2}}\paren{\A_{\ul{\theta}}\pm i\A_{\ul{\phi}}},
}
which satisfies $\A_-=\A_+^\dagger$.
We write down the KK expansions for a bulk gauge field in the Feynman-'t Hooft gauge $\xi=1$:
\al{
\A_\mu(x,\theta,\phi)
	&=	\sum_{\ell=0}^{\infty}\sum_{m=-\ell}^{\ell}
		\A_\mu^{\ell m}(x)\,f_\A^{\ell m}(\theta,\phi),  &
\A_+(x,\theta,\phi)
	&=	\sum_{\ell=1}^{\infty}\sum_{m=-\ell}^{\ell}
		\A_+^{\ell m}(x)\,f_+^{\ell m}(\theta,\phi),
		\label{gauge_boson_KK_expansion}
}
where the KK wave function is given by
\al{
f_\A^{\ell m}(\theta,\phi)
	&=	c_m{Y_{\ell m}(\theta,\phi)\over R}, &
f_+^{\ell m}(\theta,\phi)
	&=	{c_m\over\sqrt{2}}{{}_1Y_{\ell m}(\theta,\phi)\over  R},
}
with $c_0=1$ and $c_m=\sqrt{2}$ for $m\neq 0$.
From the reality condition $\A^\dagger = \A$, a negative $m$ mode vector is related to the corresponding positive $m$ mode: $\A^{\ell m}_{\mu}(x)=(-1)^{m}\A^{\dagger\, \ell,-m}_{\mu}(x)$.
The KK mass spectrum is
\al{
M_\ell
	&=	{\sqrt{\ell(\ell+1)}\over R}
}
for both $\A_\mu$ and $\A_\pm$.
We note that the extra dimensional components $\A_\pm$ do not have a zero mode and that $\ell$ starts from 1 for $\A_\pm$. 
The number of degrees of freedom is $2(2\ell+1)$ for each of $\A_\mu$ and $\A_\pm$, among which $2\ell+1$ of the $\A_\pm$ KK mode is absorbed by the corresponding $\A_\mu$ KK mode and $2\ell+1$ remain as physical scalars in the unitary gauge $\xi\to\infty$~\cite{DO_to_appear}.
When the bulk vector field gets a bulk mass $M$ from a spontaneous symmetry breaking, the resultant KK mass (in $\xi=1$ gauge) becomes~\cite{DO_to_appear}
\al{
M_\ell
	&=	\sqrt{{\ell(\ell+1)\over R^2}+M^2}.
}

Several comments are in order. The above KK wave-functions are all single-valued for $\phi\to \phi+2\pi$ and  do not have $\phi$ dependence at poles except for $\A_{\pm}$~\cite{DO_to_appear}. 
In terms of cotangent space vector, only $\A_{\ul\theta}$ has $\phi$ dependence at poles. 
We can interpret that the choice of the cotangent space basis $e^{\ul\theta}$ depends on $\phi$ in the limit $\epsilon\to+0$ for $\theta=\epsilon$ and $\theta=\pi-\epsilon$.
Since the Lagrangian is still invariant under the isometry of the sphere, the corresponding KK angular momentum, in particular the quantum number $m$, is conserved in our $RP^2$ model.
We note that all the KK wave functions are written in terms of the spin-weighted spherical harmonics and that the Lagrangian is manifestly spin-weight 0.
 
\section{Universal Extra Dimensions on Two-Sphere}
\label{UED_sphere}
We consider a UED model compactified on sphere.
We follow the particle content and 6D-chirality by Dobrescu and Poppitz~\cite{Dobrescu:2001ae}:
\al{
\mathcal{Q}_{+}, \mathcal{U}_{-}, \mathcal{D}_{-}, \mathcal{L}_{+}, \mathcal{N}_{-},\mathcal{E}_{-}.
	\label{S2_bulk_spinors}
}
We assign $U(1)_X$ charge $qn=-1$ for these fields and $q=0$ for the bulk Higgs field.
The KK wave functions and their masses are already given above.
The physics of $S^2$ UED is almost equivalent to that of the $S^2/Z_2$ model~\cite{Maru:2009wu} except that half of the massive KK modes are projected away in the latter.
There is no problem for the electroweak symmetry breaking by the bulk Higgs potential.
A massless fermion can acquire a mass through the Yukawa coupling
\al{
\mathcal{L}_{\text{Yukawa}}&=
	-y_{D}\ol{\mathcal{Q_{+}}}\mathcal H\mathcal{D}_{-}
	-y_{U}\ol{\mathcal{Q_{+}}}\epsilon \mathcal H^*\mathcal{U}_{-}
	-y_{E}\ol{\mathcal{L_{+}}}\mathcal H\mathcal{E}_{-}
	-y_{N}\ol{\mathcal{L_{+}}}\epsilon \mathcal H^*\mathcal{N}_{-}
+\text{h.c.}
}
The lightest massive KK mode becomes stable because of the KK angular momentum conservation, providing a possible dark matter candidate.\footnote{
In~\cite{Maru:2009wu}, it is argued that the $Z_2'$ parity $\paren{x,\theta,\phi}\to\paren{x,\theta,\phi+\pi}$ is essential for having a stable KK particle. From the point of view of this paper, this is the direct consequence of the KK angular momentum conservation, that is, we can directly read it off from the KK wave function written in spin-weighted spherical harmonics.
}

Let us illustrate the situation comparing with a five dimensional theory.
For the $S^1$ compactification, stability of the lightest massive KK particle is already guaranteed due to the KK number conservation. $S^1/Z_2$ was introduced to yield a chiral spinor zero mode.
In six dimensions, sphere compactification already gives us a 4D-chiral fermions and therefore our $S^2$ model is more economical than the $S^2/Z_2$ model.

So far the $S^2$ UED model seems successful.
However, there is a serious problem in this model.
From the KK reduction~\eqref{gauge_boson_KK_expansion}, we see that there remains an extra massless $U(1)_X$-gauge boson after compactification, which transmits a long-range force.
Such a long-range force is severely constrained by a torsion balance experiment~\cite{Smith:1999cr} to be $g_{X4}^2\lesssim 10^{-46}$ and hence
the KK mass scale, given by Eq.~\eqref{radius}, is constrained by $R^{-1}\lesssim \text{MeV}/n$.
(Note that the KK scale $R^{-1}$ is proportional to $g_{X4}$ and hence is constrained from above by the torsion balance experiment rather than from below as in ordinary constraints.)
Such a low KK scale contradicts with the absence of direct observation at colliders up to now. The minimal $S^2$ UED model cannot be consistent to the torsion balance experiment.\footnote{
This problem also persists in the $S^2/Z_2$ model~\cite{Maru:2009wu}, where it is assumed that the $U(1)_X$ is broken by an anomaly in order to avoid the problem.
It might be somewhat unclear how the classical monopole configuration, which is essential for the RSS spontaneous compactification, is maintained when the symmetry is broken by an anomaly at an UV cut-off scale.
}
In this Letter, we propose to utilize the projective plane in order to overcome this difficulty.

Finally let us comment on a 4D-Majorana mass for the 4D-right-handed neutrino, which is necessary for the see-saw mechanism to work successfully.
In~\cite{Dobrescu:2001ae} the SM-neutral (would-be 4D-right-handed) neutrino $\mathcal{N}_-$ is introduced to cancel pure gravitational anomaly.
In six dimensions, the 6D-charge conjugation gives the same 6D-chirality, that is, $(\Psi_\pm)^C$ still has an eigenvalue $\pm1$ for $\Gamma^{\ul 7}$ (and contains $(\psi_\pm)^c$ rather than $(\psi_\mp)^c$, where superscripts $C$ and $c$ denote the 6D and 4D charge conjugations, respectively).
Therefore, the 6D Majorana mass term: $\ol{(\mathcal{N}_-)^C}\mathcal{N}_-$ automatically vanishes and the right handed neutrino $\mathcal{N}_-$ cannot serve for the seesaw mechanism.\footnote{
Even if we introduce further field $\mathcal{N}_+$ and put a large 6D-Dirac mass term $\ol{\mathcal{N}_+}\mathcal{N}_-$, the resultant seesaw matrix yields one (still) massless neutrino and two heavy ones.
}

\section{Universal Extra Dimensions on Real Projective Plane}
\label{UED_RP2}
The real projective plane $RP^2$ is a sphere $S^2$ with its antipodal points identified by $\paren{\theta,\phi}\sim\paren{\pi-\theta,\phi+\pi}$.
Under this identification, fields can generically be transformed by a symmetry of the action.
That is, a field is not necessarily single-valued on $RP^2$ but can be multiply-defined up to a symmetry that we want to impose. 
A field must be single-valued only on a universal cover, which is $S^2$ in our case.
In particular, since $\pi_1(RP^2)=Z_2$, a field identification $\Phi(x,\theta,\phi)=\mathcal{P}\paren{\Phi(x,\pi-\theta,\phi+\pi)}$ must be by a transformation $\mathcal{P}$ that gives identity when squared: $\mathcal{P}^2=1$.

Since $RP^2$ is a non-orientable manifold, there always is a circle that gives a parity transformation when one circulates around it and comes back to the original point.
To achieve a chiral four dimensional theory after the compactification, we identify by a 6D-$CP$ transformation so that the four dimensional theory is chiral even if {\color{black} a} six dimensional theory is vector-like.

Let us start with the matter content same as in the $S^2$ UED model~\eqref{S2_bulk_spinors}, as well as
the $SU(3)_C\times SU(2)_W\times U(1)_Y$ gauge fields and the $U(1)_X$ one introduced above.
As in the ordinary UED model, the electroweak symmetry breaking $SU(2)_W\times U(1)_Y\to U(1)_\text{EM}$ is achieved by a vev of a bulk Higgs field $\mathcal H$ due to its bulk potential.

A natural antipodal identification for the $U(1)_X$ field is the 6D-$CP$ transformation that is consistent to and does not violate the $U(1)_X$ monopole configuration~\eqref{monopole_classical_configuration}:\footnote{\label{Xtheta_KK}
$C$ changes the sign of all $X_M$ and 6D-$P$ transformation further changes the sign of $X_\theta$. This $CP$ is essentially the same as the one in~\cite{Lim:2009pj}. 
Note that the even field $X_\theta$ does not have a zero mode since it picks put the curvature of $S^2$ and its KK expansion (together with $X_\phi$) starts from $\ell=1$ mode as can be seen from Eq.~\eqref{gauge_boson_KK_expansion}. See also footnote~\ref{Aphi_KK}.
}
\al{
{X_\mu(x,\pi-\theta,\phi+\pi)}^N
	&={X^C_\mu(x,\theta,\phi)}^S=	-{X_\mu(x,\theta,\phi)}^S, \nn
{X_\phi(x,\pi-\theta,\phi+\pi)}^N
	&={X^C_\phi(x,\theta,\phi)}^S=	-{X_\phi(x,\theta,\phi)}^S, \nn
{X_\theta(x,\pi-\theta,\phi+\pi)}^N
	&=-{X^C_\theta(x,\theta,\phi)}^S=	 {X_\theta(x,\theta,\phi)}^S.
}
This identification by 6D-$CP$ transformation is an outer automorphism, see~\cite{Hebecker:2001jb} for its use in five dimensions. 
We find that only odd $\ell$ modes survive under this projection.
In particular the zero mode of $U(1)_{X}$ is projected out and hence the $U(1)_{X}$ symmetry is broken by the projection.

The SM fermions must have a $U(1)_X$ charge in order to have a massless mode after KK reduction.
In order to have an invariant $U(1)_X$ interaction under this antipodal projection, the $U(1)_X$-charged SM fermions must have the following transformation property, which is nothing but a 6D-$CP$ transformation:
\al{
\Psi^N(x,\pi-\theta,\phi+\pi)
	&=	\eta_\Psi\Gamma^{\ul{5}}\paren{\Psi^{S}(x,\theta,\phi)}^{C}
	=	-\xi_\Psi\eta_\Psi\Gamma^{\ul 2}\Psi^{S*}(x,\theta,\phi), \label{spinor_z_2_transformation}
}
where $\xi_\Psi$ and $\eta_\Psi$ are the intrinsic parities for 6D-$C$ and $P$ transformations, respectively: $|\xi_\Psi|=|\eta_\Psi|=1$.
Note that  $\paren{\Psi^C}{}^C=-\Psi$ with our conventions. 
We put the 6D-$CP$ intrinsic parity $\xi_\Psi\eta_\Psi=\pm 1$ for matter~\eqref{S2_bulk_spinors} with 6D-chirality $\pm$, and accordingly $\mp 1$ for mirror matter~\eqref{6D_mirror_fermions} with $\pm$ chirality.
We use another basis $\ul{5}$ and $\ul{6}$ according to~\cite{RandjbarDaemi:1982hi}. In this basis, under the projection, the six dimensional basis $\del_{\ul A}$ transform as 
\al{
\del_{\mu}&\to \del_{\mu}, & \del_{\ul{5}}{}^{\ns}&\to -\del_{\ul{5}}{}^{\sn}, & \del_{\ul{6}}{}^{\ns}&\to \del_{\ul{6}}{}^{\sn}.
}
Hence  $\Gamma^{\ul{5}}$ in \eqref{spinor_z_2_transformation} means 6D-$P$ transformation.
 
This identification necessitates a mirror field having opposite 6D-chirality and opposite SM charges per each field in Eq.~\eqref{S2_bulk_spinors}:
\al{
\mc Q_-, \mc U_+, \mc D_+, \mc L_-, \mc N_+, \mc E_+,
\label{6D_mirror_fermions}
}
where we set $U(1)_X$ charge for the mirror fermions~\eqref{6D_mirror_fermions} to be $qn=-1$ and therefore the same as that of the ordinary one~\eqref{S2_bulk_spinors}.
We doubled the field content by adding a mirror field with opposite 6D-chirality per each field, and then these mirror fields~\eqref{6D_mirror_fermions} will be identified-away with~\eqref{S2_bulk_spinors} by the antipodal projection.

In order to realize proper SM gauge interactions between zero modes, we find that the SM gauge fields $\A$ should be identified by the 6D-$P$ transformation
\al{
\A_\mu(x,\pi-\theta,\phi+\pi)
	&=	\A_\mu(x,\theta,\phi), \nn
\A_\phi(x,\pi-\theta,\phi+\pi)
	&=	\A_\phi(x,\theta,\phi), \nn
\A_\theta(x,\pi-\theta,\phi+\pi)
	&=	-\A_\theta(x,\theta,\phi). 
}
Note that only even $\ell$ modes survive under this identification.
In the $S^2/Z_2$ orbifold model~\cite{Maru:2009wu}, half of modes in each KK level are projected away, while in our model, whole modes in each odd $\ell$ level are projected away for the SM gauge fields.
In particular there remains a zero mode for and only for $\A_\mu$.\footnote{\label{Aphi_KK}
Note that $\A_\phi$ does not allow a zero mode since it picks up the curvature of $S^2$, namely, the KK expansion~\eqref{gauge_boson_KK_expansion} starts from $\ell=1$ for the extra-dimensional component $\A_+$ even before the antipodal projection.
See also footnote~\ref{Xtheta_KK}.
}
The SM gauge interaction must be of the following 6D-chiral form:
\al{
S_\text{int}
	&=	-ig_6\int d^6x\sqrt{-g}\paren{
			\ol{\Psi_+}\Gamma^M\A_M\Psi_+
			-\ol{\Psi_-}\Gamma^M\A_M^*\Psi_-
			},
			\label{6Dchiral_gauge}
}
where $\Psi$ stands for $\mathcal Q$, $\mathcal U^C$, $\mathcal D^C$, $\mathcal L$, $\mathcal N^C$ and $\mathcal E^C$, with $C$ being the 6D-charge conjugation that does not change the 6D-chirality.
Note that both the $\Psi_+$ and $\Psi_-$ are irreducible representations of Lorentz transformation and hence are totally independent of each other at this stage.
We also note that when we redefine the chirality minus field by the 6D-charge conjugation as $\tilde\Psi_-=\Psi_-^C$, we get completely vector-like SM gauge interaction
\al{
S_\text{int}
	&=	-ig_6\int d^6x\sqrt{-g}\paren{
			\ol{\Psi_+}\Gamma^M\A_M\Psi_+
			+\ol{\tilde\Psi_-}\Gamma^M\A_M\tilde\Psi_-
			}.
}
This is why we have assigned SM charges such that 6D-chirality plus spinor $\Psi_+$ has opposite charges to original chirality minus spinor $\Psi_-$.
With this identification and charge assignments, $U(1)_{X}$ symmetry is broken by the compactification at the KK scale $\sim R^{-1}$, while all the SM gauge fields have zero modes after the compactification.

Next we discuss the electroweak symmetry breaking in our model. Since the Higgs doublet does not have $U(1)_{X}$ charge, it can be trivially identified under the antipodal projection (by a 6D-$P$ transformation):
\al{
\mathcal H(x,\pi-\theta,\phi+\pi)=\mathcal H(x,\theta,\phi). 
}
The Higgs field has expansion~\eqref{KK_wave_function_for_scalar} in terms of the ordinary spherical harmonics $Y_{\ell m}={}_0Y_{\ell m}$ and can have a vev that is constant in extra dimensions.
Then we can realize ordinal electroweak symmetry breaking and obtain $W^{\pm}$, $Z$ and $A$ as four dimensional zero modes.

\begin{figure}[tn]
\begin{center}
\includegraphics[width=0.6\textwidth]{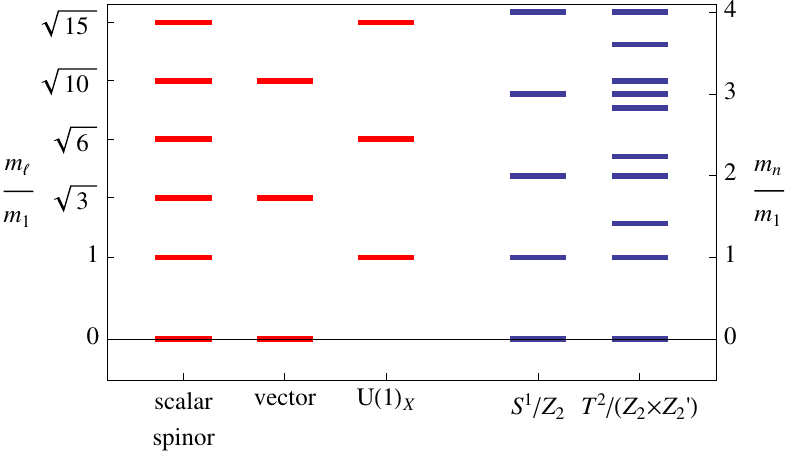}
  \end{center}
\caption{Tree level KK mass spectra in units of the 1st KK mass $m_1$, neglecting bulk masses.
\label{KK_mass_spectra}}
\end{figure}

We summarize our KK mass specta in Fig.~\ref{KK_mass_spectra}, which are distinctive from those in the 5D and 6D UED models with orbifolded toroidal compacitifications: $S^1/Z_2$ and $T^2/(Z_2\times Z_2')$, respectively.
We have shown the tree-level KK mass in units of the first KK mass $m_1$ that is the physical quantity observed at collider.\footnote{
Recall that in our model $m_1=\sqrt{2}/R$, while in the $S^1/Z_2$ and $T^2/(Z_2\times Z_2')$ models $m_1=1/R$. The KK masses are then given as $\left|n\right|m_1$ with $n\in\mathbb Z$ for the $S^1/Z_2$ model and $\sqrt{n_1^2+n_2^2}\,m_1$ with $n_1,n_2\in \mathbb Z$ for the $T^2/(Z_2\times Z_2')$ model.
}
We have plotted the KK mass up to $4m_1$ neglecting the bulk mass $M$, which should be generated by the electroweak symmetry breaking for SM fields but is negligible as long as $M^2\ll R^{-2}\simeq \text{TeV}^2$.

We introduce 6D-Dirac Yukawa interaction, e.g.\ for down-type quarks as follows:
\al{
\mathcal{\mathcal{L}}_{\text{Yukawa}}=-\sqrt{-g}\left[
	y_{\mathcal D}
		\paren{\ol{\mathcal{Q}_{+}} \mathcal H\mathcal{D}_{-}
		+\ol{\mathcal Q_-^C} \mathcal H\mathcal D_+^C}
	+\text{h.c.} \right], 
	\label{Yukawa}
}
and similarly for up-type quarks and leptons. 
Note that this 6D Yukawa interaction violates the 6D-$CP$ invariance since the Yukawa couplings include complex phases. We identify 6D spinors and $U(1)_{X}$ vector by 6D-$CP$ transformation, while identifying Higgs and SM gauge fields by 6D-$P$ transformation.  This 6D-$CP$ violating interaction allows such identifications of 6D fields and we can get effective 4D-$CP$ violating Yukawa and gauge interactions.  
Namely, under the antipodal identification, $\mathcal Q_+$ and $\mathcal Q_-^C$  as well as $\mathcal D_-$ and $\mathcal D_+^C$ are interchanged so that the sum of the terms in the parentheses in Eq.~\eqref{Yukawa} are left invariant under the identification even if it violates 6D-$CP$.

In our $RP^2$ model, we can write a following 
6D bulk mass term for and only for the SM-neutral~$\mathcal{N}_{\pm}$
\al{
\mathcal{\mathcal{L}}_{\mathcal{N}}:=-\sqrt{-g}\paren{M_{\mathcal{N}}\ol{\mathcal{N}_{+}}\mathcal{N}_{-}+M_{\mathcal{N}}^{*}\ol{\mathcal{N}_{-}}\mathcal{N}_{+}}. 
}
Note that the fermion bilinear $\ol{\Psi_+}\Psi_-$ is odd for each of the 6D-$P$ and $C$ transformations and hence is even for the 6D-$CP$ transformation. 
This  term gives 4D Majorana mass of the 4D-right-handed neutrino after the antipodal identification and the dimensional reduction. 
Our UED model on $RP^2$ overcomes the problem of appearance of the $U(1)_{X}$ zero mode, as well as the absence of the Majorana mass term of 4D-right-handed neutrino.

\section{Discussions}

First let us discuss 6D anomalies.
In the $RP^2$ model, irreducible gauge anomalies, which cannot be canceled by the Green-Schwarz mechanism~\cite{Green:1984sg,Dine:1987xk}, are
$[SU(3)_{C}]^3\,U(1)_{Y}$, $[SU(3)_{C}]^3\,U(1)_{X}$ and $[U(1)_Y]^3U(1)_X$.\footnote{
Actually we may cancel such a mixed gauge anomaly by adding a four form field whose field strength is dual to that of a scalar, see e.g.~\cite{Scrucca:2004jn}. In that case, $U(1)_X$ gauge boson will become superheavy and will spoil the RSS spontaneous compactification, as is discussed in footnote~\ref{anomaly_giving_superheavy}.
}
With the matter content~\eqref{S2_bulk_spinors},\footnote{
In this discussion, we take the bulk fermions~\eqref{S2_bulk_spinors} as physical degrees of freedom and regard that the mirror fermions are identified-away.
}
the first one automatically vanishes.\footnote{
In the models on the sphere and $S^2/Z_2$, both of these irreducible anomalies automatically vanish if one introduces $\mathcal{N}_{-}$. In~\cite{Maru:2009wu}, there is no field corresponding to $\mathcal{N_{-}}$ and the $U(1)_{X}$ is assumed to be broken by an anomaly. As mentioned above, such a breaking spoils the $U(1)_X$ monopole configuration and destroys the RSS spontaneous compactification itself. 
}
A simple remedy to cancel the last two, involving $U(1)_X$, is to introduce additional heavy 6D spinors $Q^1_+,Q^2_-, L^1_+, L^2_-$ whose SM charges are the same as quark and lepton doublets while their $U(1)_X$ charge is $+1$.\footnote{
Of course we need to add their mirrors $Q^1_-,Q^2_+, L^1_-, L^2_+$ having the same $U(1)_X$ charge $+1$ and the chiral gauge interaction~\eqref{6Dchiral_gauge} between $Q^1_+$ and $Q^1_-$ etc.
}
We note that both before and after adding these fermions, the pure gravitational anomaly automatically cancels since there are equal number of $+$ and $-$ fields.
Also, irreducible mixed gravitational anomalies: $[U(1)_{X}]^2[\text{graviton}]^2$, $[U(1)_{Y}]^2[\text{graviton}]^2$, and $U(1)_{X}U(1)_{Y}[\text{graviton}]^2$ are automatically canceled before and after adding them.
Note also that we can freely make new spinors heavy by introducing bulk mass term $M\ol{Q^1_+}Q^2_-$ etc.\footnote{
After we integrate out such new heavy spinors, Wess-Zumino-like terms~\cite{D'Hoker:1984pc} might be induced at lower energies. It would be worth studying this issue further.
}
All the remaining local anomalies can be absorbed by the Green-Schwarz mechanism~\cite{Dobrescu:2001ae,Borghini:2001sa}. 
It would be worth studying if three generations are required by the $SU(2)_W$ global anomaly, see also footnote~\ref{anomaly_footnote}.

{\color{black} In this paper we have neglected gravitational modes.}\footnote{\color{black}
A technique is given in~\cite{Parameswaran:2009bt} that may be useful for the explicit computation of the full spectrum including bulk graviton contributions.
}
Since the same isometry remains in $RP^2$ after the antipodal projection, there are several massless modes from the fluctuation of graviton, which are 4D gravitons and 4D isometry gauge field (non-abelian graviphoton).
We note that the lowest modes of the gravi-scalars are all massive of order $1/R$ and that the radion is stabilized by the RSS spontaneous compactification~\cite{Salvio:2007mb}. The existence of massless graviphoton violates the experimental constraint~\cite{Smith:1999cr}. 
However spin-weight of the isometry gauge fields is $\pm 1$~\cite{RandjbarDaemi:1982hi,Salvio:2007mb,RandjbarDaemi:1983bw,RandjbarDaemi:2006gf}.  
Therefore we suppose that these graviphotons cannot propagate between zero-mode SM matters, which have spin-weight 0 at the tree level although such an interaction might be loop induced.\footnote{
If this is not the case, we would need further orbifold projection on $RP^2$ to kill its isometry.
}

Finally let us comment on the cosmological constraint.
In the $RP^2$ model, the SM gauge bosons have only even KK modes.
In particular, the fist KK photon is projected away and cannot be a dark matter, unlike the minimal 5D UED model.
Therefore the most phenomenologically natural mass spectrum would be that the first $U(1)_X$ gauge boson is the LKP and the first KK neutrino is the Next Lightest KK Particle (NLKP) or vice versa.\footnote{
The loop corrections to the $U(1)_X$ KK masses should be suppressed by the factor $g_{X4}^2\sim 10^{-28}n^2$ in our model. We also note that there is no orbifold fixed point in our model, which is the source to the positive mass shift to the fermion KK modes in the 5D orbifold UED model~\cite{Cheng:2002iz}. Detailed study for the loop corrections is being performed. In worst case, we can still arrange the small tree-level mass splitting at the UV cutoff scale on phenomenological basis, as in Ref.~\cite{Servant:2002aq}. 
}
Since we expect that the tree level mass splitting at the UV cutoff scale would be small compared to the KK scale~\cite{Servant:2002aq}, the supposed NLKP decays into the LKP with lifetime relevant to the bound coming from neutrino injection~\cite{Kanzaki:2007pd}, if the above mass splitting is realized.
Detailed study of this issue will be presented in a separate publication.


\subsection*{Acknowledgment}
We are grateful to Satoshi Yamaguchi for helpful discussions
and thank Yutaka Hosotani, Daisuke Ida, Kazunori Kohri, Shigeki Matsumoto, Takahiro Nishinaka, Takaaki Nomura, Tetsuya Onogi, {\color{black} Marco Serone,} Noburo Shiba, and Minoru Tanaka for useful comments. 
The work of K.O.\ is partly supported by Scientific Grant by Ministry of Education and Science (Japan), Nos.\ 19740171, 20244028, and 20025004.

\bibliography{Reference}
\bibliographystyle{TitleAndArxiv}

  \end{document}